\documentclass[superscriptaddress, 
reprint,
 amsmath,amssymb,
 aps,
]{revtex4-1}

\usepackage{graphicx}
\usepackage{dcolumn}
\usepackage{bm}

\begin{document}

\preprint{APS/123-QED}

\title{Optimal Filling of Shapes}

\author{Carolyn L. Phillips}
\affiliation{Applied Physics Program, University of Michigan, Ann Arbor, Michigan, 48109, USA}
\author{Joshua A. Anderson}
\affiliation{Department of Chemical Engineering, University of Michigan, Ann Arbor, Michigan, 48109, USA}
\author{Greg Huber}
\affiliation{Richard Berlin Center for Cell Analysis \& Modeling, University of Connecticut Health Center, Farmington, CT 06062, USA}
\affiliation{Department of Cell Biology, University of Connecticut Health Center, Farmington, CT 06062, USA}
\author{Sharon C. Glotzer}
\email{Corresponding author \emph{E-mail address:} sglotzer@umich.edu}
\affiliation{Applied Physics Program, University of Michigan, Ann Arbor, Michigan, 48109, USA}
\affiliation{Department of Chemical Engineering, University of Michigan, Ann Arbor, Michigan, 48109, USA}


\begin{abstract}
We present {\it filling} as a type of spatial subdivision problem similar to covering and packing. Filling addresses the optimal placement of overlapping objects lying entirely inside an arbitrary shape so as to cover the most interior volume.  In $n$-dimensional space, if the objects are polydisperse $n$-balls, we show that solutions correspond to sets of maximal $n$-balls.  For polygons, we provide a heuristic for finding solutions of maximal discs.  We consider the properties of ideal distributions of $N$ discs as $N\rightarrow \infty$. We note an analogy with energy landscapes.
\end{abstract}

\maketitle

Packings of non-overlapping objects such as monodisperse or polydisperse spheres, ellipsoids, or polyhedra have been long-studied by physicists and mathematicians \cite{PhysRevLett.106.148302,PhysRevLett.106.135702,PhysRevLett.106.125503,PhysRevLett.106.115704,PhysRevLett.105.068001,PhysRevLett.104.185501,PhysRevLett.99.155501,PhysRevLett.96.225502,PhysRevLett.94.015502,PhysRevLett.92.044301,PhysRevE.53.2571}.  Coverings of shapes by overlapping objects are also of interest in many physical settings\cite{PhysRevE.82.056109,PhysRevD.79.104017,PhysRevLett.99.180602,PhysRevLett.72.3745,PhysRevLett.83.3986}.  Packing and covering are two familiar examples of problems in the subdivision of space subject to prescribed constraints.   Whereas in the packing problem, objects packed in a given shape are not allowed to overlap each other or the shape boundary, in the covering problem objects overlap both each other and the shape boundary in an effort to maximally cover a given shape. In both problems, the objects may be monodisperse or polydisperse in size. In covering, the objects are typically $n$\emph{-balls} (discs, in $2D$); in packing, the objects may be of any shape.   

\begin{figure}[h!]
\includegraphics{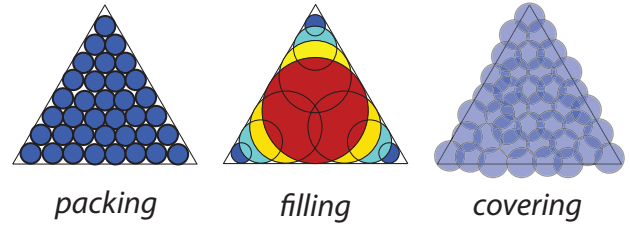}
\caption{\label{fig:packcoverfill} The problem of filling a shape, such as a triangle, may be viewed as intermediate between the familiar packing and covering problems.}
\end{figure}

In this Letter, we present a new type of spatial subdivision problem which can be viewed as intermediate between the packing and covering problems.  We define \emph{filling} as the problem of packing overlapping objects inside of a defined shape so as to cover the interior volume without extending beyond the boundary of the shape (Fig.~\ref{fig:packcoverfill}).  We are primarily interested in the optimal filling of an $n$-dimensional shape, characterized by a well-defined $n-1$ dimensional surface, with $N$ polydisperse $n$-balls.  Specifically, we seek the optimal placement and radii of the $n$-balls for a given $N$. 

The notion of filling arises from the problem of modeling anisotropic nanoparticles as rigid bodies composed of a sum of isotropic volume-excluding potentials\cite{PhysRevLett.95.056105, *citeulike:1065652}. Other applications are the problem of irradiating a tumor with the fewest number of beam shots, while controlling the beam diameter, but without damaging surrounding tissue\cite{Bourland}; using time-delayed sources to create shaped wavefronts; combining precision-placed explosives with tunable blast radii; positioning proximity sensors with defined radii; cell phone and wireless network coverage; or any problem of ablation or deposition where one has a sharp impenetrable boundary and a radially tunable tool.  It also can be related to the coarsening (due to Ostwald ripening) of wet foams packed in containers with non-wetting surfaces. 

Here we show that, to optimally fill an $n$-dimensional shape with $N$ $n$-balls of varying radii, only solutions of \emph{maximal} $n$-balls (that is, balls whose centers lie on the medial axis of the shape) need be considered. It will follow that the dimension of the solution space is reduced from $n+1$ to $n-1$. Secondly, we consider optimal fillings of polygons with $N$ polydisperse overlapping discs.  For this two-dimensonal case, we present a heuristic for the numerical generation of optimal solutions for arbitrary $N$.  Thirdly, by considering how discs are optimally distributed in polygons as $N\rightarrow \infty$, we derive exact analytical expressions for the spatial distribution of the discs and show that the fraction of unfilled area for optimal solutions vanishes like $1/N^2$.  The analytical expressions may be used to approximate solutions for finite but large $N$. We derive an exact expression for the fractional allocation of discs over the three medial axis branches of a triangle as $N\rightarrow \infty$.  We discuss how solutions for $n=2$ provide insight into the filling problem of generalized shapes in arbitrary dimensions.  We also note an interesting connection to energy landscapes. 
  
\noindent\emph{Reducing the dimension of the solution space using maximal $n$-balls.}  
For optimal filling solutions, the objective function to be maximized is defined to be the volume of the union of  a set of $N$ $n$-balls constrained to the interior of a shape $G$. The upper bound of this function is the total volume of $G$.  The optimal $N=1$ filling is the largest $n$-ball that can fit in $G$.  Given a set of $N$ $n$-balls, the contribution of a single ball to the total filling is equal to the volume of the ball minus the fractional share of the volume of any overlap with the $N-1$ other balls (SI, \S1). To find optimal solutions for $G$, we find it is not necessary to consider the space of all possible $n$-balls contained in $G$.   Importantly, we show that only the maximal $n$-balls need be considered.

First introduced by Blum\cite{blum1967,blum1978} as a ``topological skeleton'', the medial axis is a reduction of an $n$-dimensional shape into an $n-1$-dimensional space, $M(G)$, the locus of centers of the maximal $n$-balls.    A \emph{maximal $n$-ball} is an $n$-ball completely contained in the shape tangent to the shape boundary at two or more points.  Also, a maximal $n$-ball is a ball contained completely in $G$ but not contained in any other ball in $G$.

The radius function $r$ is a continuous, non-negative function defined at each point of $M(G)$ as the radius of the maximal $n$-ball centered at that point.  The medial axis and the radius function comprise a complete shape descriptor\cite{blum1978} and can be used to reconstruct $G$.   Every maximal $n$-ball in $G$ can be represented as a unique point, its center, on $M(G)$.  If each $n$-ball in a proposed optimal filling is replaced by a maximal $n$-ball containing it, the new solution will fill an equivalent or greater area of $G$.   Therefore, only solutions of maximal $n$-balls need be generated. Finding solutions is reduced to finding center points on an $n-1$-dimensional hypersurface.

\begin{figure*}
\includegraphics{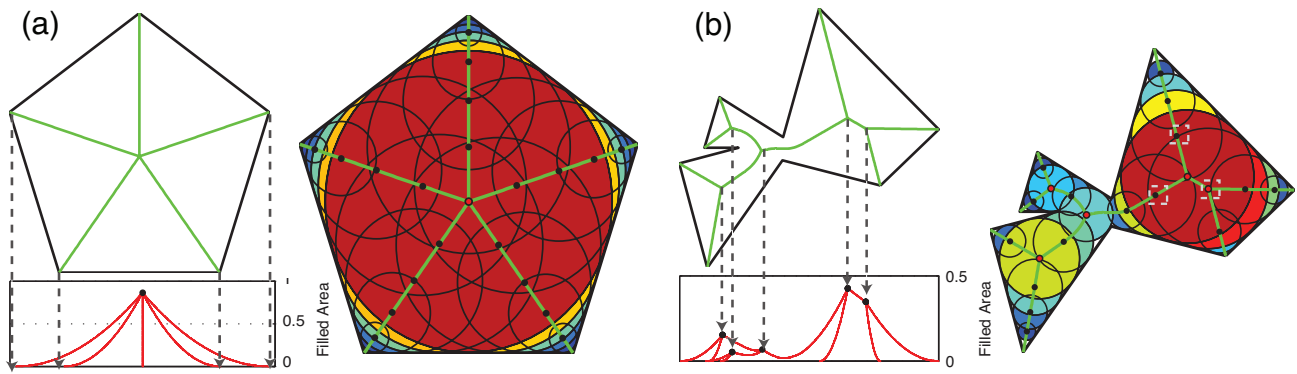}
\caption{\label{fig:landscape_pentagon_shape19} Two polygons and their medial axes (green line segments):
(a) pentagon and (b) concave polygon.  Below each is a side view of the landscape of the $N=1$ total filled area function (the area of both shapes is normalized to 1).  For the pentagon, the horizontal extents of the shape are shown by dashed lines.  For the concave shape, the local filling maxima of the $N=1$ solution are shown by dashed lines. To the right of each shape is an optimal filling with 21 discs.  Disc centers in traps are shown as red open dots.  For the concave polygon, dashed squares enclose the centers of neighbors of the largest maximal disc.}
\label{landscape}
\end{figure*}

\noindent\emph{Heuristic for 2D shapes.}
For a planar shape $G$, $M(G)$ is the one-dimensional planar graph that is the locus of the centers of the maximal discs (2-balls) of the shape.  $M(G)$ is a set of 1-manifolds, or branches, plus connecting branch points and terminating end points\cite{blum1978}.   For a convex polygon, $M(G)$ is composed only of straight segments.  For polygons\footnote{In this paper, we mean \emph{simple} polygons, i.e. convex or concave, but not self-intersecting.}, $M(G)$ is composed of straight segments and, possibly, parabolic curves.   Various algorithms exist to compute the medial axes of convex or concave polygons\cite{Vilaplana,preparata}.   The medial axes of a pentagon and a concave polygon are shown in Fig.~\ref{fig:landscape_pentagon_shape19}.  The filled area of all the $N=1$ maximal disc fillings is shown from the side in Fig.~\ref{fig:landscape_pentagon_shape19} for both shapes.  The global maximum is synonymous with the largest disc inscribable in $G$.  In a convex shape, there is only one maximum.  In a concave shape, there can be many. 

The \emph{neighbors} of a disc $A$ in $M(G)$ are the set of centers that can be reached along any path in $M(G)$ originating at the center of $A$ without traversing another center.  In $2D$, we find that the change in the total filling due to locally displacing a maximal disc center on $M(G)$ is a function of only the change in the overlaps of the maximal disc with its neighbors(SI, \S1).  In Fig.~\ref{fig:landscape_pentagon_shape19}, the centers of the neighbors of the largest disc of the concave polygon are enclosed by dashed boxes.   

\emph{Traps} are a special set of points on the medial axis, which are important in optimal filling solutions because they are often occupied by centers. To see this, we consider the behavior of the objective function around a trap. We begin by observing that the filling function is piece-wise first-order continuous, with fixed points of first-order discontinuity that we refer to as junctions.  These are points where the radius function and/or path in $M(G)$ is discontinuous.   In $2D$ all branch points are junctions.  If a junction is a local maximum with respect to moving a single disc center (all other disc centers held fixed), then the junction is a trap, because small displacements in the centers of discs neighboring the discontinuity do not displace the position of the local maximum.  In Fig.~\ref{fig:landscape_pentagon_shape19}, the centers caught in traps in the optimal filling solutions are shown as red open dots.  We observe that disc centers in traps tend to be common features in optimal filling solutions, and even a fixed feature when $N$ is large.   

\begin{figure}[b]
\includegraphics{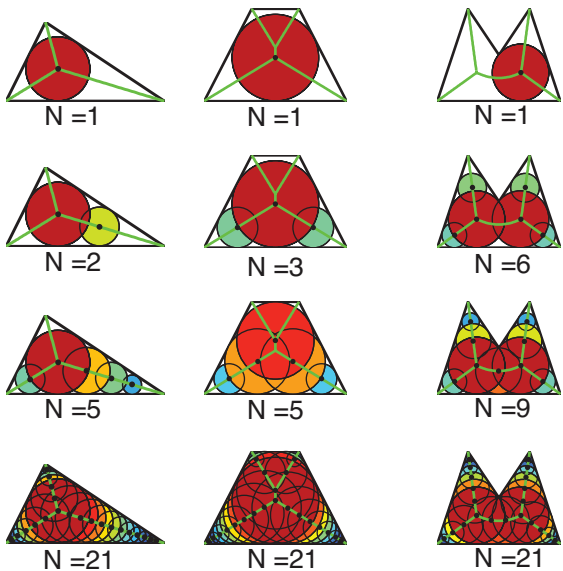}
\caption{\label{fig:table} Maximal filling solutions for polygons}
\end{figure}

We now propose a solution strategy whereby disc centers are distributed onto $M(G)$ and the local filling maximum is found by a gradient method.  If the maxima of enough distribution samplings are generated such that all the local maxima can be enumerated, then the global filling maximum is among them.  We propose the following heuristic that accomplishes this efficiently while also greedily using the $N-1$ solution to intelligently reduce the number of distributions to be searched:
(1)  The medial axis is divided into K pieces, branches with monotonically increasing radius functions and the junctions connecting them.
(2)  It is assumed that there is at most one local maximum per \emph{way}, $W$, of partitioning the $N$ discs over the K pieces (branches and junctions), $W= \{n_i\}_1^{K}$ where $N=\sum_i^K n_i$  and $n_i$ is the number of discs on the $ith$ piece, $n_i \in \mathbb{N} $. 
(3)  It is further assumed that, given the optimal way of partitioning $N-1$ discs, $\{n'_i\}_1^{K}$, the optimal way of partitioning $N$ discs is nearby, where nearby means $\sum_1^K \mid n_i - n'_i \mid$ is small, and that if the discs assigned to a given piece are decreased in number, the pieces that have discs increased in number have a minimal distance (counted by number of connecting pieces) to the decreased piece.
(4)  The local maxima of the nearby ways are generated using any local maximum finding technique (e.g. active set or sequential quadratic programming optimization schemes\cite{nocedal06}).  The best local maximum found is presumed to be the optimal $N$ filling solution for the shape.  Note that during the local maximum search, the searched space is first-order continuous because the fixed points of first-order discontinuities are in the set of junctions.  Non-fixed points of first-order discontinuity that correspond to the point of tangency of discs in $G$ are not generally relevant because they cannot be points of local maximum.

This heuristic is made more efficient by taking advantage of center-occupied traps and the dependence of the filling function on the nearest neighbors.   When a trap is occupied by a center in the $N-1$ solution, the phase space of centers can be divided into independent sub-spaces.  If it is known (or guessed) that the best solution for $N$ also includes a center in the trap, then the sub-parts of $M(G)$ connected only by the center-occupied trap can be searched independently.  Rearrangements of centers in one sub-part cannot affect the best arrangement of centers in another if they are connected only by the center-occupied trap.  

We implemented this heuristic for polygons, which have only a few types of medial axis branches.  The sets of discs shown in Figs.~\ref{fig:landscape_pentagon_shape19} and \ref{fig:table} are the best solutions found by the heuristic and verified by a genetic algorithm\cite{dcgmanuscript}.  The heuristic generally produces a superior solution to the genetic algorithm, as long as a large enough neighborhood of ways is considered. One surprising finding is that the largest disc that fits in $G$, i.e. the $N=1$ solution, is not always part of the solution for $N>1$.  In Fig. \ref{fig:table}, for example, the $N=5$ solution for the trapezoid does not include the $N$=1 solution.

\noindent\emph{Optimally filling a polygon as N $\rightarrow \infty$}.  It is instructive to examine how the optimal filling of a shape converges to the total volume of the shape as $N\rightarrow \infty$.   We find that this limit can be solved exactly for polygons.  The $M(G)$ of a polygon can be divided into branches of only three types, (1) straight segments with linear radius functions, (2) parabolic curves with quadratic radius functions, and (3) straight segments with square root radius functions (SI, \S2).  Immediately, case (3) can be ignored because no disc center on such a curve fills more area than that filled by two discs placed at the ends of the curve.  In ideal solutions, Case (3) type curves are empty except for the ends of the curve.  Thus we need only derive an expression for the relative density of centers as a function of position on the other two branch types.  Along each branch type, the distribution of optimal filling solutions approaches this expression as $N\rightarrow \infty$.  

Let $\rho(t)$ represent the density of centers, $r(t)$ the radius function, $\kappa(t)$ the local curvature, and $(x(t), y(t))$
the position along a parameterized branch of $M(G)$, with $t \in [t_a,t_b]$.  Given an expression for the unfilled area $A_i$ along the path $i$ of the form 
\begin{eqnarray}
A_i = \textstyle\int^{t_b}_{t_a} C_i(\kappa, r', r)\frac{dt}{\rho^2},
\label{eq:genarea}
\end{eqnarray}
where $C_i$ is a function to be determined, we wish to determine the function $\rho(t)$ that minimizes this area constrained by
$
N =  \textstyle\int^{t_b}_{t_a} \rho dt.
\label{eq:fixedN}
$
Note that if we sum the unfilled areas $A_i$ over all of $M(G)$, then the filled area is  $A_{G} - \sum A_i$ , where $A_{G}$ is the area of G.  
This variational problem can be solved by constructing the Lagrangian
\begin{equation}
\mathcal{L}[\rho(t); \lambda] =  \textstyle\int^{t_b}_{t_a} \left(C_i(t)\frac{1}{\rho^2} - \lambda \rho \right) dt
\end{equation}
and taking the pointwise derivative with respect to $\rho(t)$,
\begin{equation}
\frac{\partial \mathcal{L}}{\partial \rho(\tau)} =  \textstyle\int^{t_b}_{t_a} \left( \frac{-2C_i(t)}{\rho^3} +\frac{\partial}{\rho^2\partial \rho} C_i(t) -  \lambda  \right) \delta (t-\tau)dt = 0.
\end{equation}
This equation is solved by functions $\rho$ that satisfy
\begin{equation}
\textstyle
-2C_i(t) +\rho \frac{\partial}{\partial \rho} C_i(t) -  \rho^3\lambda  = 0
\label{eq:lagrange}
\end{equation}
Note that $ \rho =  \left(\frac{C_i(t)}{\lambda}\right)^{1/3}
\label{lagrangesolution}$
 satisfies this equation.  
 
For Case (1), where $(x(t),y(t)) = \bf{A}$$t +\bf{B}$,  and $r(t) = ct + r_0$, for constants $c,r_0 \geq 0$, where $t_a \geq 0$, we derive (SI, \S3) 
\begin{eqnarray}
C= {\textstyle\frac{1}{12r}\left(1-r'^2\right)^{3/2}}.
\label{eq:nocurve}
\end{eqnarray}
It follows that $\rho \propto r^{-1/3}$.

For Case (2),  where $(x(t),y(t)) = (2r_0t,r_0t^2)$, $r = r_0(1 + t^2)$, where $r_0$ is the minimum of the radius function, and $\kappa(t) = \textstyle\frac{1}{2r_0}\ \left(1+t^2\right)^{-3/2}$, we derive (SI, \S4)
\begin{equation}\textstyle
C =  \frac{1}{12}\left(\frac{r_0\kappa}{r}\right)
=   \frac{1}{24r_0}\left({1+t^2}\right)^{-5/2}.
\end{equation}  
It follows that  $\rho \propto  \left({1+t^2}\right)^{-5/6} \propto r^{-5/6}$.

For both Case (1) and Case (2), the distribution of centers follows a power law with respect to the local radius function.  We observe that centers on $M(G)$ will be distributed more densely where the radius function is smaller.   
Given $\rho = \rho_0r^{-\alpha}$, for $\alpha$ = 1/3 or 5/6,  $\rho_0$ can be determined from
\begin{eqnarray}
 \rho_0=&N \left(\int^{t_a}_{t_b}  r^{-\alpha} dt\right)^{-1} =N/R_0.
\end{eqnarray}
$R_0$ is then a constant determined by the radius function of the branch.
  
For Case (1), we observe that the distribution of centers on the medial axis branch is also \emph{scale-free}.   The distribution of centers follows a power law with respect to the distance from the vertex (where $t=0$) of the polygon.
Eq.~\ref{eq:genarea} becomes
\begin{eqnarray} \textstyle
A = \frac{1}{N^2}\int^{t_b}_{t_a} R_0^2 C(\kappa, r', r)dt =  \frac{1}{N^2} \mathcal{C}
\end{eqnarray}
\noindent where $\mathcal{C}$ is the evaluated integral.
Thus, as $N\rightarrow \infty$ for a system of ideally distributed centers, the filling converges to the area of $G$ with an asymptotic error proportional to $N^{-2}$.  It can be presumed that all shapes that can be approximated by polygons with an increasing number of sides also converge with an $N^{-2}$ error term. 

If we divide $M(G)$ into $k$ branches we can predict the partitioning of the discs over the branches as $N\rightarrow\infty$.    The fraction of discs on a given branch $i$ is (see SI, \S5)
\begin{equation}
f_i = \frac{(\mathcal{C}_i)^{1/3}}{ (\mathcal{C}_1)^{1/3} + (\mathcal{C}_2)^{1/3} + \cdots + (\mathcal{C}_k)^{1/3}}. \label{gendist}
\end{equation}

For a triangle, which is composed of three Case (1) branches, the fraction of the discs on a given branch can be solved analytically:

\begin{equation}
f_i =  \frac{\textrm{cot}(\theta_i)}{ \textrm{cot}(\theta_1)+ \textrm{cot}(\theta_2) +\textrm{cot}(\theta_3)} \label{triangledist}
\end{equation}

\noindent where $\theta_1, \theta_2$ and $\theta_3$ are the internal angles of the triangle, each of which is associated with a branch of the medial axis.  From Eq.~\ref{triangledist}, it is clear that the optimal solution preferentially populates medial axis branches associated with smaller internal angles.  This can be observed in the triangles in column one of Fig.~\ref{fig:table}.

\noindent{\emph Discussion.} A convenient framework for visualizing filling solutions of a $2D$ shape is to consider the unfilled area as the ``energy'' of the system, and the force acting on a disc center as the negative gradient of this energy.  The force acting on a single center can be divided into two parts, a force due only to the local radius function, and a purely repulsive short range force (also dependent on the radius function) between a center and its neighbors (SI, \S1C).   The range is defined by where two discs overlap.  Because this energy function is not first-order continuous, these force definitions have discontinuities.  A center in a trap, therefore, has a restoring force in each path direction away from the trap.  A center not in a trap is at a point where the local forces balance.  The filled area plots of Fig.~\ref{fig:landscape_pentagon_shape19}, therefore, is like an inverted energy landscape of the $N$ = 1 system of the two polygons.

In this work we defined the filling problem for arbitrary dimensions. Although we examined the details of its solution structure only in two spatial dimensions, we can extrapolate some of our findings to higher dimensions. For instance, in the case of polygons, we found that first order continuous manifolds meet at lower dimension manifolds where centers are trapped. It is natural to expect similar behavior for $D>2$.  Further, higher dimensional polyhedron shapes are also likely to have manifolds with scale-free solutions.  Such behavior will be investigated in future publications.

\emph{Acknowledgements.}
We acknowledge E. Chen for reviewing our mathematics, Suresh Krishnan for software help and A. Haji-Akbar, M. Engel and T.J. Ligocki for interesting discussions.  C.L.P acknowledges the U.S. DOE CSGF program.  S.C.G and C.L.P acknowledge the U.S. DOE DE- FG02-02ER46000.  S.C.G and J.A.A acknowledge the DOD under Award No. N00244-09-1-0062.  

\bibliography{prl}

\begin{thebibliography}{26}%
\makeatletter
\providecommand \@ifxundefined [1]{%
 \@ifx{#1\undefined}
}%
\providecommand \@ifnum [1]{%
 \ifnum #1\expandafter \@firstoftwo
 \else \expandafter \@secondoftwo
 \fi
}%
\providecommand \@ifx [1]{%
 \ifx #1\expandafter \@firstoftwo
 \else \expandafter \@secondoftwo
 \fi
}%
\providecommand \natexlab [1]{#1}%
\providecommand \enquote  [1]{``#1''}%
\providecommand \bibnamefont  [1]{#1}%
\providecommand \bibfnamefont [1]{#1}%
\providecommand \citenamefont [1]{#1}%
\providecommand \href@noop [0]{\@secondoftwo}%
\providecommand \href [0]{\begingroup \@sanitize@url \@href}%
\providecommand \@href[1]{\@@startlink{#1}\@@href}%
\providecommand \@@href[1]{\endgroup#1\@@endlink}%
\providecommand \@sanitize@url [0]{\catcode `\\12\catcode `\$12\catcode
  `\&12\catcode `\#12\catcode `\^12\catcode `\_12\catcode `\%12\relax}%
\providecommand \@@startlink[1]{}%
\providecommand \@@endlink[0]{}%
\providecommand \url  [0]{\begingroup\@sanitize@url \@url }%
\providecommand \@url [1]{\endgroup\@href {#1}{\urlprefix }}%
\providecommand \urlprefix  [0]{URL }%
\providecommand \Eprint [0]{\href }%
\@ifxundefined \urlstyle {%
  \providecommand \doi  [0]{\begingroup \@sanitize@url \@doi}%
  \providecommand \@doi [1]{\endgroup \@@startlink {\doibase
  #1}doi:\discretionary {}{}{}#1\@@endlink }%
}{%
  \providecommand \doi  [0]{doi:\discretionary{}{}{}\begingroup
  \urlstyle{rm}\Url }%
}%
\providecommand \doibase [0]{http://dx.doi.org/}%
\providecommand \Doi [0]{\begingroup \@sanitize@url \@Doi }%
\providecommand \@Doi  [1]{\endgroup\@@startlink{\doibase#1}\@@Doi}%
\providecommand \@@Doi [1]{#1\@@endlink}%
\providecommand \selectlanguage [0]{\@gobble}%
\providecommand \bibinfo  [0]{\@secondoftwo}%
\providecommand \bibfield  [0]{\@secondoftwo}%
\providecommand \translation [1]{[#1]}%
\providecommand \BibitemOpen [0]{}%
\providecommand \bibitemStop [0]{}%
\providecommand \bibitemNoStop [0]{.\EOS\space}%
\providecommand \EOS [0]{\spacefactor3000\relax}%
\providecommand \BibitemShut  [1]{\csname bibitem#1\endcsname}%
\bibitem [{\citenamefont {Lespiat}\ \emph {et~al.}(2011)\citenamefont
  {Lespiat}, \citenamefont {Cohen-Addad},\ and\ \citenamefont
  {H\"ohler}}]{PhysRevLett.106.148302}%
  \BibitemOpen
  \bibfield  {author} {\bibinfo {author} {\bibfnamefont {R.}~\bibnamefont
  {Lespiat}}, \bibinfo {author} {\bibfnamefont {S.}~\bibnamefont
  {Cohen-Addad}}, \ and\ \bibinfo {author} {\bibfnamefont {R.}~\bibnamefont
  {H\"ohler}},\ }\Doi {10.1103/PhysRevLett.106.148302} {\bibfield  {journal}
  {\bibinfo  {journal} {PRL},\ }\textbf {\bibinfo {volume} {106}},\ \bibinfo
  {pages} {148302} (\bibinfo {year} {2011})}\BibitemShut {NoStop}%
\bibitem [{\citenamefont {Jacquin}\ \emph {et~al.}(2011)\citenamefont
  {Jacquin}, \citenamefont {Berthier},\ and\ \citenamefont
  {Zamponi}}]{PhysRevLett.106.135702}%
  \BibitemOpen
  \bibfield  {author} {\bibinfo {author} {\bibfnamefont {H.}~\bibnamefont
  {Jacquin}}, \bibinfo {author} {\bibfnamefont {L.}~\bibnamefont {Berthier}}, \
  and\ \bibinfo {author} {\bibfnamefont {F.}~\bibnamefont {Zamponi}},\ }\Doi
  {10.1103/PhysRevLett.106.135702} {\bibfield  {journal} {\bibinfo  {journal}
  {PRL},\ }\textbf {\bibinfo {volume} {106}},\ \bibinfo {pages} {135702}
  (\bibinfo {year} {2011})}\BibitemShut {NoStop}%
\bibitem [{\citenamefont {Zhao}\ \emph {et~al.}(2011)\citenamefont {Zhao},
  \citenamefont {Tian},\ and\ \citenamefont {Xu}}]{PhysRevLett.106.125503}%
  \BibitemOpen
  \bibfield  {author} {\bibinfo {author} {\bibfnamefont {C.}~\bibnamefont
  {Zhao}}, \bibinfo {author} {\bibfnamefont {K.}~\bibnamefont {Tian}}, \ and\
  \bibinfo {author} {\bibfnamefont {N.}~\bibnamefont {Xu}},\ }\Doi
  {10.1103/PhysRevLett.106.125503} {\bibfield  {journal} {\bibinfo  {journal}
  {PRL},\ }\textbf {\bibinfo {volume} {106}},\ \bibinfo {pages} {125503}
  (\bibinfo {year} {2011})}\BibitemShut {NoStop}%
\bibitem [{\citenamefont {Mughal}\ \emph {et~al.}(2011)\citenamefont {Mughal},
  \citenamefont {Chan},\ and\ \citenamefont {Weaire}}]{PhysRevLett.106.115704}%
  \BibitemOpen
  \bibfield  {author} {\bibinfo {author} {\bibfnamefont {A.}~\bibnamefont
  {Mughal}}, \bibinfo {author} {\bibfnamefont {H.~K.}\ \bibnamefont {Chan}}, \
  and\ \bibinfo {author} {\bibfnamefont {D.}~\bibnamefont {Weaire}},\ }\Doi
  {10.1103/PhysRevLett.106.115704} {\bibfield  {journal} {\bibinfo  {journal}
  {PRL},\ }\textbf {\bibinfo {volume} {106}},\ \bibinfo {pages} {115704}
  (\bibinfo {year} {2011})}\BibitemShut {NoStop}%
\bibitem [{\citenamefont {Hoy}\ and\ \citenamefont
  {O'Hern}(2010)}]{PhysRevLett.105.068001}%
  \BibitemOpen
  \bibfield  {author} {\bibinfo {author} {\bibfnamefont {R.~S.}\ \bibnamefont
  {Hoy}}\ and\ \bibinfo {author} {\bibfnamefont {C.~S.}\ \bibnamefont
  {O'Hern}},\ }\Doi {10.1103/PhysRevLett.105.068001} {\bibfield  {journal}
  {\bibinfo  {journal} {PRL},\ }\textbf {\bibinfo {volume} {105}},\ \bibinfo
  {pages} {068001} (\bibinfo {year} {2010})}\BibitemShut {NoStop}%
\bibitem [{\citenamefont {Jaoshvili}\ \emph {et~al.}(2010)\citenamefont
  {Jaoshvili}, \citenamefont {Esakia}, \citenamefont {Porrati},\ and\
  \citenamefont {Chaikin}}]{PhysRevLett.104.185501}%
  \BibitemOpen
  \bibfield  {author} {\bibinfo {author} {\bibfnamefont {A.}~\bibnamefont
  {Jaoshvili}}, \bibinfo {author} {\bibfnamefont {A.}~\bibnamefont {Esakia}},
  \bibinfo {author} {\bibfnamefont {M.}~\bibnamefont {Porrati}}, \ and\
  \bibinfo {author} {\bibfnamefont {P.~M.}\ \bibnamefont {Chaikin}},\ }\Doi
  {10.1103/PhysRevLett.104.185501} {\bibfield  {journal} {\bibinfo  {journal}
  {PRL},\ }\textbf {\bibinfo {volume} {104}},\ \bibinfo {pages} {185501}
  (\bibinfo {year} {2010})}\BibitemShut {NoStop}%
\bibitem [{\citenamefont {Kamien}\ and\ \citenamefont
  {Liu}(2007)}]{PhysRevLett.99.155501}%
  \BibitemOpen
  \bibfield  {author} {\bibinfo {author} {\bibfnamefont {R.~D.}\ \bibnamefont
  {Kamien}}\ and\ \bibinfo {author} {\bibfnamefont {A.~J.}\ \bibnamefont
  {Liu}},\ }\Doi {10.1103/PhysRevLett.99.155501} {\bibfield  {journal}
  {\bibinfo  {journal} {PRL},\ }\textbf {\bibinfo {volume} {99}},\ \bibinfo
  {pages} {155501} (\bibinfo {year} {2007})}\BibitemShut {NoStop}%
\bibitem [{\citenamefont {Donev}\ \emph {et~al.}(2006)\citenamefont {Donev},
  \citenamefont {Stillinger},\ and\ \citenamefont
  {Torquato}}]{PhysRevLett.96.225502}%
  \BibitemOpen
  \bibfield  {author} {\bibinfo {author} {\bibfnamefont {A.}~\bibnamefont
  {Donev}}, \bibinfo {author} {\bibfnamefont {F.~H.}\ \bibnamefont
  {Stillinger}}, \ and\ \bibinfo {author} {\bibfnamefont {S.}~\bibnamefont
  {Torquato}},\ }\Doi {10.1103/PhysRevLett.96.225502} {\bibfield  {journal}
  {\bibinfo  {journal} {PRL},\ }\textbf {\bibinfo {volume} {96}},\ \bibinfo
  {pages} {225502} (\bibinfo {year} {2006})}\BibitemShut {NoStop}%
\bibitem [{\citenamefont {Radin}\ and\ \citenamefont
  {Sadun}(2005)}]{PhysRevLett.94.015502}%
  \BibitemOpen
  \bibfield  {author} {\bibinfo {author} {\bibfnamefont {C.}~\bibnamefont
  {Radin}}\ and\ \bibinfo {author} {\bibfnamefont {L.}~\bibnamefont {Sadun}},\
  }\Doi {10.1103/PhysRevLett.94.015502} {\bibfield  {journal} {\bibinfo
  {journal} {PRL},\ }\textbf {\bibinfo {volume} {94}},\ \bibinfo {pages}
  {015502} (\bibinfo {year} {2005})}\BibitemShut {NoStop}%
\bibitem [{\citenamefont {Baram}\ \emph {et~al.}(2004)\citenamefont {Baram},
  \citenamefont {Herrmann},\ and\ \citenamefont
  {Rivier}}]{PhysRevLett.92.044301}%
  \BibitemOpen
  \bibfield  {author} {\bibinfo {author} {\bibfnamefont {R.~M.}\ \bibnamefont
  {Baram}}, \bibinfo {author} {\bibfnamefont {H.~J.}\ \bibnamefont {Herrmann}},
  \ and\ \bibinfo {author} {\bibfnamefont {N.}~\bibnamefont {Rivier}},\ }\Doi
  {10.1103/PhysRevLett.92.044301} {\bibfield  {journal} {\bibinfo  {journal}
  {PRL},\ }\textbf {\bibinfo {volume} {92}},\ \bibinfo {pages} {044301}
  (\bibinfo {year} {2004})}\BibitemShut {NoStop}%
\bibitem [{\citenamefont {Aste}(1996)}]{PhysRevE.53.2571}%
  \BibitemOpen
  \bibfield  {author} {\bibinfo {author} {\bibfnamefont {T.}~\bibnamefont
  {Aste}},\ }\Doi {10.1103/PhysRevE.53.2571} {\bibfield  {journal} {\bibinfo
  {journal} {Phys. Rev. E},\ }\textbf {\bibinfo {volume} {53}},\ \bibinfo
  {pages} {2571} (\bibinfo {year} {1996})}\BibitemShut {NoStop}%
\bibitem [{\citenamefont {Torquato}(2010)}]{PhysRevE.82.056109}%
  \BibitemOpen
  \bibfield  {author} {\bibinfo {author} {\bibfnamefont {S.}~\bibnamefont
  {Torquato}},\ }\Doi {10.1103/PhysRevE.82.056109} {\bibfield  {journal}
  {\bibinfo  {journal} {Phys. Rev. E},\ }\textbf {\bibinfo {volume} {82}},\
  \bibinfo {pages} {056109} (\bibinfo {year} {2010})}\BibitemShut {NoStop}%
\bibitem [{\citenamefont {Messenger}\ \emph {et~al.}(2009)\citenamefont
  {Messenger}, \citenamefont {Prix},\ and\ \citenamefont
  {Papa}}]{PhysRevD.79.104017}%
  \BibitemOpen
  \bibfield  {author} {\bibinfo {author} {\bibfnamefont {C.}~\bibnamefont
  {Messenger}}, \bibinfo {author} {\bibfnamefont {R.}~\bibnamefont {Prix}}, \
  and\ \bibinfo {author} {\bibfnamefont {M.~A.}\ \bibnamefont {Papa}},\ }\Doi
  {10.1103/PhysRevD.79.104017} {\bibfield  {journal} {\bibinfo  {journal}
  {Phys. Rev. D},\ }\textbf {\bibinfo {volume} {79}},\ \bibinfo {pages}
  {104017} (\bibinfo {year} {2009})}\BibitemShut {NoStop}%
\bibitem [{\citenamefont {Anteneodo}\ and\ \citenamefont
  {Morgado}(2007)}]{PhysRevLett.99.180602}%
  \BibitemOpen
  \bibfield  {author} {\bibinfo {author} {\bibfnamefont {C.}~\bibnamefont
  {Anteneodo}}\ and\ \bibinfo {author} {\bibfnamefont {W.~A.~M.}\ \bibnamefont
  {Morgado}},\ }\Doi {10.1103/PhysRevLett.99.180602} {\bibfield  {journal}
  {\bibinfo  {journal} {PRL},\ }\textbf {\bibinfo {volume} {99}},\ \bibinfo
  {pages} {180602} (\bibinfo {year} {2007})}\BibitemShut {NoStop}%
\bibitem [{\citenamefont {Coutinho}\ \emph {et~al.}(1994)\citenamefont
  {Coutinho}, \citenamefont {Coutinho-Filho}, \citenamefont {Gomes},\ and\
  \citenamefont {Nemirovsky}}]{PhysRevLett.72.3745}%
  \BibitemOpen
  \bibfield  {author} {\bibinfo {author} {\bibfnamefont {K.~R.}\ \bibnamefont
  {Coutinho}}, \bibinfo {author} {\bibfnamefont {M.~D.}\ \bibnamefont
  {Coutinho-Filho}}, \bibinfo {author} {\bibfnamefont {M.~A.~F.}\ \bibnamefont
  {Gomes}}, \ and\ \bibinfo {author} {\bibfnamefont {A.~M.}\ \bibnamefont
  {Nemirovsky}},\ }\Doi {10.1103/PhysRevLett.72.3745} {\bibfield  {journal}
  {\bibinfo  {journal} {PRL},\ }\textbf {\bibinfo {volume} {72}},\ \bibinfo
  {pages} {3745} (\bibinfo {year} {1994})}\BibitemShut {NoStop}%
\bibitem [{\citenamefont {Verberkmoes}\ and\ \citenamefont
  {Nienhuis}(1999)}]{PhysRevLett.83.3986}%
  \BibitemOpen
  \bibfield  {author} {\bibinfo {author} {\bibfnamefont {A.}~\bibnamefont
  {Verberkmoes}}\ and\ \bibinfo {author} {\bibfnamefont {B.}~\bibnamefont
  {Nienhuis}},\ }\Doi {10.1103/PhysRevLett.83.3986} {\bibfield  {journal}
  {\bibinfo  {journal} {PRL},\ }\textbf {\bibinfo {volume} {83}},\ \bibinfo
  {pages} {3986} (\bibinfo {year} {1999})}\BibitemShut {NoStop}%
\bibitem [{\citenamefont {Horsch}\ \emph {et~al.}(2005)\citenamefont {Horsch},
  \citenamefont {Zhang},\ and\ \citenamefont
  {Glotzer}}]{PhysRevLett.95.056105}%
  \BibitemOpen
  \bibfield  {author} {\bibinfo {author} {\bibfnamefont {M.~A.}\ \bibnamefont
  {Horsch}}, \bibinfo {author} {\bibfnamefont {Z.}~\bibnamefont {Zhang}}, \
  and\ \bibinfo {author} {\bibfnamefont {S.~C.}\ \bibnamefont {Glotzer}},\
  }\Doi {10.1103/PhysRevLett.95.056105} {\bibfield  {journal} {\bibinfo
  {journal} {PRL},\ }\textbf {\bibinfo {volume} {95}},\ \bibinfo {pages}
  {056105} (\bibinfo {year} {2005})}\BibitemShut {NoStop}%
\bibitem [{\citenamefont {Chen}\ \emph {et~al.}(2007)\citenamefont {Chen},
  \citenamefont {Zhang},\ and\ \citenamefont {Glotzer}}]{citeulike:1065652}%
  \BibitemOpen
  \bibfield  {author} {\bibinfo {author} {\bibfnamefont {T.}~\bibnamefont
  {Chen}}, \bibinfo {author} {\bibfnamefont {Z.}~\bibnamefont {Zhang}}, \ and\
  \bibinfo {author} {\bibfnamefont {S.~C.}\ \bibnamefont {Glotzer}},\ }\href
  {http://dx.doi.org/10.1073/pnas.0604239104} {\bibfield  {journal} {\bibinfo
  {journal} {Proc Natl Acad Sci},\ }\textbf {\bibinfo {volume} {104}},\
  \bibinfo {pages} {717} (\bibinfo {year} {2007})}\BibitemShut {NoStop}%
\bibitem [{\citenamefont {Bourland}\ and\ \citenamefont {Wu}(1996)}]{Bourland}%
  \BibitemOpen
  \bibfield  {author} {\bibinfo {author} {\bibfnamefont {J.~D.}\ \bibnamefont
  {Bourland}}\ and\ \bibinfo {author} {\bibfnamefont {Q.~R.}\ \bibnamefont
  {Wu}},\ }in\ \href {http://portal.acm.org/citation.cfm?id=647241.718952}
  {\emph {\bibinfo {booktitle} {Proc. of the 4th Inter. Conf. on Visualization
  in Biomedical Computing}}}\ (\bibinfo  {publisher} {Springer-Verlag},\
  \bibinfo {address} {London, UK},\ \bibinfo {year} {1996})\ pp.\ \bibinfo
  {pages} {553--558}\BibitemShut {NoStop}%
\bibitem [{\citenamefont {Blum}(1967)}]{blum1967}%
  \BibitemOpen
  \bibfield  {author} {\bibinfo {author} {\bibfnamefont {H.}~\bibnamefont
  {Blum}},\ }\href
  {http://pageperso.lif.univ-mrs.fr/\~{}edouard.thiel/rech/1967-blum.pdf}
  {\bibfield  {journal} {\bibinfo  {journal} {Models for the Perception of
  Speech and Visual Form},\ \bibinfo {pages} {362}} (\bibinfo {year}
  {1967})}\BibitemShut {NoStop}%
\bibitem [{\citenamefont {Blum}\ and\ \citenamefont {Nagel}(1978)}]{blum1978}%
  \BibitemOpen
  \bibfield  {author} {\bibinfo {author} {\bibfnamefont {H.}~\bibnamefont
  {Blum}}\ and\ \bibinfo {author} {\bibfnamefont {R.~N.}\ \bibnamefont
  {Nagel}},\ }\Doi {10.1016/0031-3203(78)90025-0} {\bibfield  {journal}
  {\bibinfo  {journal} {Pattern Recognition},\ }\textbf {\bibinfo {volume}
  {10}},\ \bibinfo {pages} {167} (\bibinfo {year} {1978})}\BibitemShut
  {NoStop}%
\bibitem [{Note1()}]{Note1}%
  \BibitemOpen
  \bibinfo {note} {In this paper, we mean \protect \emph {simple} polygons,
  i.e. convex or concave, but not self-intersecting.}\BibitemShut {Stop}%
\bibitem [{\citenamefont {Vilaplana}(1996)}]{Vilaplana}%
  \BibitemOpen
  \bibfield  {author} {\bibinfo {author} {\bibfnamefont {J.}~\bibnamefont
  {Vilaplana}},\ }\href@noop {} {\enquote {\bibinfo {title} {Computing the
  medial axis transform of polygonal objects by testing discs},}\ } (\bibinfo
  {year} {1996})\BibitemShut {NoStop}%
\bibitem [{\citenamefont {Preparata}(1977)}]{preparata}%
  \BibitemOpen
  \bibfield  {author} {\bibinfo {author} {\bibfnamefont {F.~P.}\ \bibnamefont
  {Preparata}},\ }\href@noop {} {\bibfield  {journal} {\bibinfo  {journal}
  {Proc. 6th Symp. Math. Foundations of Comput. Sci.},\ \bibinfo {pages} {443}}
  (\bibinfo {year} {1977})}\BibitemShut {NoStop}%
\bibitem [{\citenamefont {Nocedal}\ and\ \citenamefont
  {Wright}(2006)}]{nocedal06}%
  \BibitemOpen
  \bibfield  {author} {\bibinfo {author} {\bibfnamefont {J.}~\bibnamefont
  {Nocedal}}\ and\ \bibinfo {author} {\bibfnamefont {S.~J.}\ \bibnamefont
  {Wright}},\ }\href@noop {} {\emph {\bibinfo {title} {Numerical
  Optimization}}}\ (\bibinfo  {publisher} {Springer},\ \bibinfo {year}
  {2006})\BibitemShut {NoStop}%
\bibitem [{\citenamefont {Phillips}\ \emph {et~al.}(2011)\citenamefont
  {Phillips}, \citenamefont {Anderson}, \citenamefont {Chen},\ and\
  \citenamefont {Glotzer}}]{dcgmanuscript}%
  \BibitemOpen
  \bibfield  {author} {\bibinfo {author} {\bibfnamefont {C.~L.}\ \bibnamefont
  {Phillips}}, \bibinfo {author} {\bibfnamefont {J.~A.}\ \bibnamefont
  {Anderson}}, \bibinfo {author} {\bibfnamefont {E.}~\bibnamefont {Chen}}, \
  and\ \bibinfo {author} {\bibfnamefont {S.~C.}\ \bibnamefont {Glotzer}},\
  }\href@noop {} {\bibfield  {journal} {\bibinfo  {journal} {Preprint}}
  (\bibinfo {year} {2011})}\BibitemShut {NoStop}%
\end{thebibliography}%

\end{document}